\documentclass[a4paper]{article}
\usepackage{amsmath}
\usepackage{latexsym}
\usepackage{graphicx}
\usepackage{bm}

\title{Lagrangian formulism of elasticity with relevance to surface energy}

\author{Zaixing Huang\\[5pt]
  State Key Laboratory of Mechanics and Control of Mechanical Structures\\
  Nanjing University of Aeronautics and Astronautics\\
  Yudao Street 29, Nanjing, 210016, P R China\\ E-mail: huangzx@nuaa.edu.cn}

\begin{document}

\maketitle

\begin{abstract}
By introducing the divergence of a vector potential into the
Lagrangian, a Lagrangian framework is developed to incorporate the
surface energy into elasticity. Besides the Euler-Lagrange equation
and natural boundary condition, a new boundary constitutive equation
is derived from the variation of the Lagrangian and configuration on
which the Lagrangian is defined. On the boundary surface, explicit
expression of the vector potential with respect to field variable
and surface curvature is determined. Based on this framework, an
elastic model with relevance to the surface energy is established.
The Young-Laplace's formula is generalized into elastic solid in a
new form. Making use of this model, we investigate the surface
energy effect in the radial vibration of spherical nanoparticle.
Numerical calculation shows that natural frequencies of nanoparticle
will shift down
due to the surface energy. This shift is especially apparent in the vibration of soft matter nanoparticles.\\
\textbf{Key words:} Euler-Lagrange equation, surface energy,
generalize Young-Laplace' formula, natural frequency, nanoparticle
\end{abstract}

\section{Introduction} \label{intro} From the view of atomic level,
the free surface of a bulk material is different from its interior.
This difference gives rise to a excess energy to appear on surface.
It is called the surface free energy (or shortly called the surface
energy). In classical continuum mechanics, the influence of surface
energy on the deformation of solid is usually ignored, because atoms
on surface is far smaller in magnitude than that in the interior of
bulk. However, with entering from microscale into nanoscale, the
ratio of atoms on surface to total atoms becomes gradually
significant. therefore, the surface energy effect has to be taken
into account in the deformation of solid.

On the surface energy effect in elastic deformation, early works can
be traced back to Shuttleworth \cite{1}, Herring \cite{2}, Nicolson
\cite{3} and Vermaak et al \cite{4}. They investigated the lattice
contraction caused by the surface tension in small crystal
particles. A general theoretical framework incorporating surface
energy into continuum mechanics was proposed in the 1970s by Gurtin
and Murdoch \cite{5,6}. Within this framework, Steigmann and Ogden
\cite{7} generalized the conception of surface energy to the
curvature dependence case. Huang and Wang et al \cite{8,9} clarified
the necessity to introduce three configurations in an accurate and
complete description for the deformation of continuum with surface
energy. Altenbach et al \cite{10} gave a proof on existence and
uniqueness of the solutions of initial and boundary value problems
of linear elasticity with surface energy. Schiavone and Ru \cite{11}
discussed the solvability of boundary value problems in a theory of
plane-strain elasticity concerned with surface energy.

Various models based on the Gurtin-Murdoch theory have been
presented in several contexts, e.g. nanoparticles, wires and films
(Huang and Thomson et al \cite{12}; Cammarata \cite{12,13};
Dingreville and Qu et al \cite{15}; Park \cite{16}), nanoscale rods,
beams, plates and shells (Miller and Shenoy \cite{17}; Huang
\cite{18}; Bar and Altus et al \cite{19}; Liu and Rajapakse
\cite{20}; Wang and Feng \cite{21}; Wang and Zhao et al \cite{22}),
nanoinclusions (Duan and Wang et al \cite{23}; He and Li \cite{24};
Sharma and Ganti \cite{25}). So far, a considerable number of
literatures have been devoted to the continuum models considering
surface energy. It is not the purpose of this paper to list and
review these abundant literatures. The reader can be referred to the
recent reviews by Wang and Huang et al \cite{26} and Duan, Wang and
Karihaloo \cite{27} on the literatures.

The subject of the paper will focus on developing a Lagrangian
formulation corresponding to the Gurtin-Murdoch theory. The novel
feature of this formulation consists in that a vector potential is
introduced to characterize the surface energy of continuum.
Consequently, the Eular-Lagrange equation, natural boundary
condition and a new boundary constitutive equation can be
consistently determined by simultaneously taking the variation of
the Lagrangian and configuration of continuum on which the
Lagrangian is defined.

The paper is outlined as follows.In section 2,  based on a vector
potential, a Lagrangian framework is developed to incorporate
surface energy into continuum mechanics. The boundary constitutive
equation and the vector potential are determined on the boundary
surface. By linearization, in section 3 we put forward to an elastic
model with relevance to surface energy in which a generalized
Young-Laplace's formula is derived consistently. Making use of this
elastic model, we investigate the surface energy effect in the
radial vibration of spherical nanoparticle. some interesting results
are predicted. Finally, we close this paper with a summary.
\\

\textbf{Notation:} A compact notation is used, with boldface letters
being vectors or tensors. The index rules and summation convention
are adopted. Latin indices have the range 1, 2, 3. Partial
derivatives with respect to coordinates are represented as
$\partial_k=\partial /\partial x^k$ or $(\cdot)_{,k}=\partial(\cdot)
/\partial x^k$. Partial derivative with respect to time is denoted
by an upper dot, e.g., $\dot{a}=\partial a/\partial t$. Other
symbols will be introduced in the text where they appear for the
first time.
\section{Theoretical framework}
\label{sec:1} Let $\textbf{x}=\{x^j\}$ ($j$ = 1, 2, 3) be a
3-dimensional position vector on $\Omega\subset R^3$. The Greece
letter $\Omega$ stands for current configuration of a continuum. We
use $\bm{\varphi}=\bm{\varphi}(t, \textbf{x})$ to characterize the
field defined on $[t_0, t_1]\cup\Omega$, where $t\in[t_0, t_1]$
denotes time. Depending on circumstances, $\bm{\varphi}$ is a
scalar, vector or tensor.

As is well known, if two Lagrangians differ by the divergence of a
vector function only of the field variables and coordinates, their
Euler-Lagrange equations are the same in form. Thus, if such a
divergence term is added to a Lagrangian, dynamic behaviors of field
are unchanged under the fixed boundary condition. However, if the
boundary is free, the additive divergence will contribute to the
field by entering the natural boundary condition. Therefore, a
problem worthy of being asked is whether a divergence term added to
the Lagrangian can be used to characterize the surface effects of a
free continuum or not. Motivated by this idea, we assume a
Lagrangian can be written as $\L=L(\bm{\varphi},\dot{\bm{\varphi},}
\partial_j\bm{\varphi})-\partial_k Y^k(\textbf{x}, \bm{\varphi})$,
where $Y^k=Y^k(\textbf{x}, \bm{\varphi})$ denotes a vector
potential. As thus, the action functional of $\bm{\varphi}$ can be
represented as
\begin{equation}\label{1}A[\bm{\varphi}]=\int_{t_0}^{t_1}\int_\Omega[L(\bm{\varphi}, \dot{\bm{\varphi}}, \partial_j\bm{\varphi})
-\partial_kY^k(\textbf{x}, \bm{\varphi})]\mathrm{d}v(\textbf{x})\mathrm{d}t.\end{equation}                              
In terms of the divergence theorem, $A[\bm{\varphi}]$ can be also
written as
\begin{equation}\label{2}A[\bm{\varphi}]=\int_{t_0}^{t_1}\int_\Omega L(\bm{\varphi}, \dot{\bm{\varphi}}, \partial_j\bm{\varphi})
\mathrm{d}v(\textbf{x})\mathrm{d}t-\int_{t_0}^{t_1}\int_{\partial\Omega}Y^k(\textbf{x}, \bm{\varphi})n_k\mathrm{d}a(\textbf{x})\mathrm{d}t,\end{equation}                                     
where $n_k$ denotes the unit normal vector on the boundary
$\partial\Omega$ of $\Omega$. It is necessary to emphasize the
distinction between the vector potential and surface energy. The
former is defined on continuum, including its interior and surface;
but the latter is only on the surface. In the following, we will
demonstrate that they are equivalent on surface.\\

If surface energy is concerned, the changes of the configuration of
continuum will become a factor having to be considered.
Consequently, the calculation for the variation of $A[\bm{\varphi}]$
should not only include the variation of the Lagrangian but also the
variation of the surface area and volume of continuum. Under this
consideration, the variation of $A[\bm{\varphi}]$ can be calculated
as follows:
\begin{eqnarray}\label{3}\delta A[\bm{\varphi}]&=&\int_{t_0}^{t_1}\int_\Omega\delta L\mathrm{d}v(\textbf{x})\mathrm{d}t+
\int_{t_0}^{t_1}\int_{\delta\Omega}
L\mathrm{d}v(\textbf{x})\mathrm{d}t\nonumber\\
&\qquad&-\int_{t_0}^{t_1}\int_{\partial\Omega}\delta
Y^kn_k\mathrm{d}a(\textbf{x})\mathrm{d}t
-\int_{t_0}^{t_1}\int_{\delta(\partial\Omega)}Y^kn_k\mathrm{d}a(\textbf{x})\mathrm{d}t\nonumber\\
&=&\int_{t_0}^{t_1}\int_\Omega[\frac{\partial L}{\partial
\bm{\varphi}}\delta \bm{\varphi}+ \frac{\partial
L}{\partial\dot{\bm{\varphi}}}\delta\dot{\bm{\varphi}}+\frac{\partial
L}{\partial(\partial_j\bm{\varphi})}\delta(\partial_j\bm{\varphi})]\mathrm{d}v(\textbf{x})\mathrm{d}t\nonumber\\
&\qquad&-\int_{t_0}^{t_1}\int_{\partial\Omega}\frac{\partial
Y^k}{\partial\bm{\varphi}}n_k\delta\bm{\varphi}\mathrm{d}a(\textbf{x})\mathrm{d}t+\int_{t_0}^{t_1}\int_{\delta\Omega}
L\mathrm{d}v(\textbf{x})\mathrm{d}t-
\int_{t_0}^{t_1}\int_{\delta(\partial\Omega)}Y^kn_k\mathrm{d}a(\textbf{x})\mathrm{d}t\nonumber\\
&=&\int_{t_0}^{t_1}\int_\Omega[\frac{\partial L}{\partial
\bm{\varphi}}-\frac{\mathrm{d}}{\mathrm{d}t}(\frac{\partial
L}{\partial\dot{\bm{\varphi}}})-\partial_j(\frac{\partial
L}{\partial(\partial_j\bm{\varphi})})]\delta\bm{\varphi}\mathrm{d}v(\textbf{x})\mathrm{d}t
+\int_\Omega[\frac{\partial
L}{\partial\dot{\bm{\varphi}}}\delta\bm{\varphi}]_{t_0}^{t_1}\mathrm{d}v(\textbf{x})\nonumber\\
&\qquad&+\int_{t_0}^{t_1}\int_{\partial\Omega}[\frac{\partial
L}{\partial(\partial_k\bm{\varphi})}-\frac{\partial Y^k}{\partial
\bm{\varphi}}]n_k\delta\bm{\varphi}\mathrm{d}a(\textbf{x})\mathrm{d}t\nonumber\\
&\qquad&+\int_{t_0}^{t_1}\int_{\delta\Omega}
L\mathrm{d}v(\textbf{x})\mathrm{d}t-
\int_{t_0}^{t_1}\int_{\delta(\partial\Omega)}Y^kn_k\mathrm{d}a(\textbf{x})\mathrm{d}t,\end{eqnarray}                                   
in which we have supposed that $\bm{\varphi}(\textbf{x})$,
$L(\bm{\varphi},\dot{\bm{\varphi}},
\partial_j\bm{\varphi})$ and $Y^\nu(\textbf{x}, \bm{\varphi})$ are suitably smooth functions. Ou-Yang and Helfrich \cite{28} have proved that
\begin{equation}\label{4}\int_{\delta\Omega}L\mathrm{d}v(\textbf{x})=\int_{\partial\Omega}
Ln_j\delta u^j\mathrm{d}a(\textbf{x}).\end{equation}                                                    
\begin{equation}\label{5}\int_{\delta(\partial\Omega)}Y^kn_k\mathrm{d}a(\textbf{x})=
\int_{\partial\Omega}2Hg_{kj}Y^k\delta u^j\mathrm{d}a(\textbf{x}),\end{equation}                        
where $\delta u^j$ is the variation of displacement of the area
element $\mathrm{d}a(\textbf{x})$ on the boundary surface
$\partial\Omega$, $H$ the mean curvature at \textbf{x} on
$\partial\Omega$ and $g_{kj}$ the materic tensor.
$\delta\bm{\varphi}=0$ at the two end points of time, so
substituting Eq. (\ref{4}) and (\ref{5}) into (\ref{3}) yields
\begin{eqnarray}\label{6}\delta A[\bm{\varphi}]&=&\int_{t_0}^{t_1}\int_\Omega[\frac{\partial L}{\partial
\bm{\varphi}}-\frac{\mathrm{d}}{\mathrm{d}t}(\frac{\partial
L}{\partial\dot{\bm{\varphi}}})-\partial_j(\frac{\partial
L}{\partial(\partial_j\bm{\varphi})})]\delta\bm{\varphi}\mathrm{d}v(\textbf{x})\mathrm{d}t
+\int_{t_0}^{t_1}\int_{\partial\Omega}[\frac{\partial
L}{\partial(\partial_k\bm{\varphi})}-\frac{\partial Y^k}{\partial
\bm{\varphi}}]n_k\delta\bm{\varphi}\mathrm{d}a(\textbf{x})\mathrm{d}t\nonumber\\&\qquad&\quad
+\int_{t_0}^{t_1}\int_{\partial\Omega}(Ln_j-2Hg_{kj}Y^k)\delta u^j\mathrm{d}a(\textbf{x})\mathrm{d}t.\end{eqnarray}       
According to the Hamilton's principle, $\delta A[\bm{\varphi}]=0$.
Therefore, the fundamental lemma of
variation leads to the below results:\\
Euler-Lagrange equation:
\begin{equation}\label{7}\frac{\mathrm{d}}{\mathrm{d}t}(\frac{\partial
L}{\partial\dot{\bm{\varphi}}})+\partial_k[\frac{\partial
L}{\partial(\partial_k\bm{\varphi})}]-\frac {\partial
L}{\partial\bm{\varphi}}=0.\end{equation}                         
Natural boundary condition:
\begin{equation}\label{8}\left. [\frac{\partial
L}{\partial(\partial_k\bm{\varphi})}-\frac{\partial Y^k}{\partial
\bm{\varphi}}]n_k
\right|_{\partial\Omega}=0.\end{equation}                                        
Boundary constitutive equation :
\begin{equation}\label{9}\left. (Ln_j-2Hg_{kj}Y^k)
\right|_{\partial\Omega}=0.\end{equation}                                        
Eq. (\ref{7}) and (\ref{8}) show that a divergence term, only
relevant to the field variable, added to the Lagrangian has no
effects on the Euler-Lagrange equation, but it contributes to the
natural boundary condition of field. Since $Y^k$ depends only on
$\bm{\varphi}$, $H$ in Eq. (\ref{9}) is necessarily relevant to
$\partial_\nu\bm{\varphi}$. Or else the derivative of $L$ with
respect to $\partial_k\bm{\varphi}$ will be zero, in terms of Eq.
(\ref{9}). That contradicts with Eq. (\ref{8}). Making use of Eq.
(\ref{9}), we have
\begin{equation}\label{10}\left.[\frac{\partial L}{\partial(\partial_j\bm{\varphi})}n_j-
2g_{kj}Y^k\frac{\partial H}{\partial(\partial_j\bm{\varphi})}]
\right|_{\partial\Omega}=0.\end{equation}                                        
Comparing Eq. (\ref{10}) with (\ref{8}), one can see that $H$ is
linearly dependent of $\partial_\nu\bm{\varphi}$. Subtracting Eq.
(\ref{9}) from (\ref{10}) gives
\begin{equation}\label{11}\left.[\frac{\partial
Y^k}{\partial \bm{\varphi}}n_k- 2g_{kj}Y^k\frac{\partial
H}{\partial(\partial_j\bm{\varphi})}]
\right|_{\partial\Omega}=0.\end{equation}                                        
Let $n^j$ denote the contravariant component of the unit normal
vector on $\partial\Omega$. So $n^jn_k=\delta^j_k$. By this
identity, Eq. (\ref{11}) can be rewritten as
\begin{equation}\label{12}\left.[\frac{\partial\Gamma}{\partial \bm{\varphi}}- \textbf{S}\Gamma]
\right|_{\partial\Omega}=0,\end{equation}                                        
where
\begin{equation}\label{13}\Gamma=Y^kn_k, \quad \textbf{S}=2n_k\frac{\partial
H}{\partial(\partial_k\bm{\varphi})}.\end{equation}                                        
On $\partial\Omega$, the solution of Eq. (\ref{12}) can be
represented as
\begin{equation}\label{14}\Gamma=\gamma \exp(\textbf{S}\bm{\varphi}),\end{equation}                           
where $\gamma$ is an integral constant. Inserting Eq. (\ref{13}) in
(\ref{14}) leads to
\begin{equation}\label{15}Y^kn_k=\gamma\exp[2\bm{\varphi}n_k\frac{\partial
H}{\partial(\partial_k\bm{\varphi})}],\end{equation}                                           
which characterizes the mathematical form of $Y^kn_k$ on the free
surface of $\Omega$. Eq. (\ref{15}) shows that, on surface, $Y^kn_k$
(i.e., $\Gamma$) is a surface curvature-dependent energy. So it is
just the surface energy. It should be noticed that, due to
concerning the curvature of surface, Eq. (\ref{15}) holds only on
the surface $\partial\Omega$, but not valid in the interior of
$\Omega$.

\section{Linear elastic model with relevance to surface energy}
\label{sec:3} Consider a free elastic body concerned with surface
energy. The Lagrangian of it is assumed to take the following form
(In the following, all indices are written as subscripts without
distinguishing the superscripts and subscripts.):
\begin{equation}\label{16}\L=L-Y_{k,k}=\frac{1}{2}\rho\dot{u}_k\dot{u}_k
-\frac{1}{2}C_{ijkl}u_{i,j}u_{k,l}-Y_{k,k}.\end{equation}                                         
where $u_k$ denotes the elastic displacement field. $\rho$ and
$C_{ijkl}$ are the mass density and elastic tensor. Based on the
viewpoint of Huang and Wang \cite{8}, $Y_{k,k}$ can be regarded as
elastic energy transformed from surface energy relative to the
so-called fictitious stress-free configuration.

In terms of Eq. (\ref{14}), $Y_kn_k$ is given by
\begin{equation}\label{17}Y_kn_k=\Gamma=\gamma\exp(S_ku_k).\end{equation}                                
Under the small displacement condition, Eq. (\ref{17}) is expanded
to the second order term of $u_k$, i.e.,
\begin{equation}\label{18}Y_kn_k=\gamma(1+S_ku_k+\frac{1}{2}S_kS_ju_ku_j),\end{equation}                 
where $S_k$ is determined by Eq. (\ref{13})$_2$. Just mentioned
above, the mean curvature $H$ in Eq. (\ref{13})$_2$ is linearly
dependent of $u_{k,i}$. As thus, $H$ is only represented as
$H=H_0(1+u_{k,k})$ in terms of the invariance of $H$ \cite{29}.
Here, $H_0$ is the initial mean curvature. Making use of Eq.
(\ref{13})$_2$, we have $S_k=2n_kH_0$. With help of this result, Eq.
(\ref{18}) becomes
\begin{equation}\label{19}Y_kn_k=\gamma(1+2H_0u_kn_k+2H_0^2u_kn_ku_jn_j).\end{equation}                
By means of Eq. (\ref{16}) and (\ref{19}), Eq. (\ref{7}) and
(\ref{8}) can be written as
\begin{equation}\label{20}C_{ijkl}u_{k,jl}=\rho \ddot{u}_i,\quad \text{in}\ \Omega.\end{equation}                                
\begin{equation}\label{21}C_{ijkl}u_{k,l}n_j-2\gamma H_0n_i(1+2H_0u_kn_k)=0,\quad \text{on}\ \partial\Omega.\end{equation}         
Eq. (\ref{20}) and (\ref{21}) are the equation of motion and
boundary condition to characterize the free elastic body concerned
with surface tension, respectively. By means of the Hooke's law
$\sigma_{ij}=C_{ijkl}u_{k,l}$, Eq. (\ref{21}) is rewritten as
\begin{equation}\label{22}\sigma_{ij}n_j=2\gamma H_0n_i(1+2H_0u_kn_k).\end{equation}         
If the term $2H_0u_kn_k$ is dropped, Eq. (\ref{22}) will reduce to
the Young-Laplace's formula that describes the surface pressure of
fluid. Therefore, Eq. (\ref{22}) can be regarded as a generalization
of the Young-Laplace's formula in solid, whereas $\gamma$ represents
the specific surface energy factor.

\section{Application to the radial vibration of nanoparticle}
\subsection{Characteristic frequency equation}
As an example, we consider the radial free vibration of a spherical
nanoparticle with the radius of $r_0$. Let material be isotropic.
The equations governing this vibration can be determined by Eq.
(\ref{20}) and (\ref{21}) together with $H_0=-1/r_0$ and
$C_{ijkl}=\lambda\delta_{ij}\delta_{kl}+\mu(\delta_{ik}\delta_{jl}+\delta_{il}\delta_{jk})$.
In a spherical coordinate system, they are written as
\begin{equation}\label{23}\frac{\partial^2u_r}{\partial
r^2}+\frac{2}{r}\frac{\partial u_r}{\partial
r}-\frac{2}{r^2}u_r=\frac{1}{c^2}\frac{\partial^2u_r}{\partial
t^2}.\end{equation}                                                     
\begin{equation}\label{24}[(\lambda+2\mu)\frac{\partial u_r}{\partial
r}+2(\lambda-\frac{2\gamma}{r})\frac{u_r}{r}]_{r=r_0}=-\frac{2\gamma}{r_0},\end{equation}                  
where $\lambda$ is the Lam\'{e} constant and $\mu$ the shear
modulus. $c=\sqrt{(\lambda+2\mu)/\rho}$, being the wave velocity.
Assume that the solution of Eq. (\ref{23}) subjected to (\ref{24})
to have the form below:
\begin{equation}\label{25}u_r(r, t)=\bar{u}_r+g(r)e^{i\omega\,t},\end{equation}    
where $\bar{u}_r$ reads
\begin{equation}\label{sp1}\bar{u}_r=-\frac{2\gamma}{(3\lambda+2\mu)r_0-4\gamma}r,\end{equation}            
which is the radial contraction displacement caused by the surface
enrgy \cite{12}. It should be noted that $\bar{u}_r>0$ or
$\bar{u}_r\rightarrow\infty$ when $r_0\leq
r_c=4\gamma/(3\lambda+2\mu)$. That is impossible in physics because
it indicates that the surface enrgy will cause the radial extension
of a spherical particle. Therefore, the elastic model is no longer
valid for the case that $r_0\leq r_c$. Inserting Eq. (\ref{25}) in
(\ref{23}) and (\ref{24}) leads to
\begin{equation}\label{26}\frac{\partial^2g}{\partial
r^2}+\frac{2}{r}\frac{\partial g}{\partial
r}+(\frac{\omega^2}{c^2}-\frac{2}{r^2})g=0.\end{equation}                                                     
\begin{equation}\label{27}[(\lambda+2\mu)\frac{\partial g}{\partial
r}+2(\lambda-\frac{2\gamma}{r})\frac{g}{r}]_{r=r_0}=0.\end{equation}                  
Let $k=\omega/c$ and $g(r)=f(r)/\sqrt{r}$. By means of these
substitution, Eq. (\ref{26}) can be transformed into a standard
Bessel equation. Therefore, the general solution to Eq. (\ref{27})
can be given as follows:
\begin{equation}\label{28}g(r)=Ar^{-1/2}J_{3/2}(kr)+Br^{-1/2}Y_{3/2}(kr),\end{equation}             
where $J_{3/2}(kr)$ and $Y_{3/2}(kr)$ are first kind and second kind
Bessel function of order 3/2, respectively. $A$ and $B$ are two
undetermined constants. Noticing $g(0)=0$, we have $B=0$. Thus, Eq.
(\ref{28}) reduces to
\begin{equation}\label{29}g(r)=Ar^{-1/2}J_{3/2}(kr).\end{equation}                                   
Substituting Eq. (\ref{29}) into (\ref{27}) yields
\begin{equation}\label{30}kJ\,'_{3/2}(kr_0)+\frac{(3\lambda-2\mu)r_0-8\gamma}{2(\lambda+
2\mu)r_0^2}J_{3/2}(kr_0)=0,\end{equation}                                                               
where $J\,'_{3/2}$ denotes the derivative of $J_{3/2}$ with respect
to its argument. Eq. (\ref{30}) represents the characteristic
frequency equation of the radial vibration of the spherical
nanoparticle involving the surface energy effect.

\subsection{Results and discussion}
Using Eq. (\ref{30}), we calculate the natural frequencies of two
different nanoparticles. The first is the silica gel particle, and
the second is the palladium (Pd) particle. The silica gel is a soft
matter. We take its $E=2.14$MPa, $\nu=0.48$, $\rho=1200$kg/m$^3$ and
$\gamma=22.0$mN/m. In terms of the transformation
$\lambda=E\nu/[(1+\nu)(1-2\nu)]$ and $\mu=E/[2(1+2\nu)]$, we have
$\lambda=17.35$MPa and $\mu=0.72$MPa. As thus,
$r_c=4\gamma/(3\lambda+2\mu)=1.65$nm. When calculating, the range of
$r_0$ should be greater than 1.65nm.

\begin{figure}[h]
\centering
\includegraphics{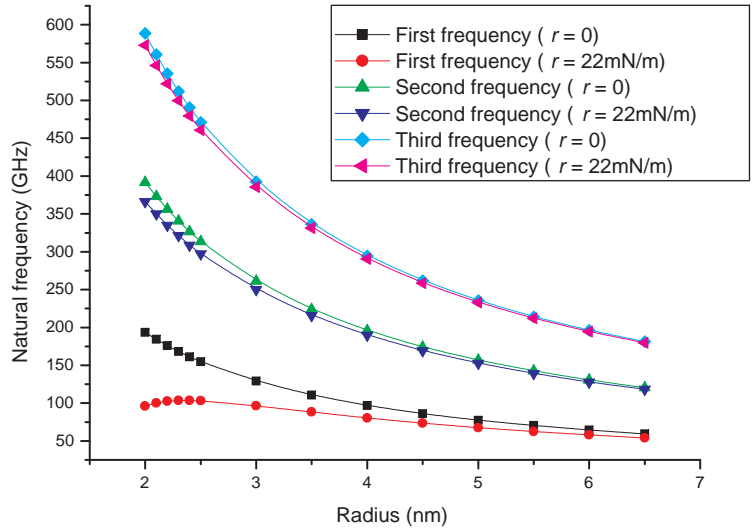} \caption{The changes of first three
natural frequencies with the radius of the silica gel
nanoparticle\label{fig:1}}
\end{figure}

The changes of first three natural frequencies with the radius $r_0$
are shown in Figure \ref{fig:1}. It can be seen that the three
natural frequencies increase with a decrease in the radius. This is
a well known size effect. More significantly, Figure 1 illustrates
that, with entering the nano-level, the influence of surface energy
on the natural frequency becomes apparent gradually. The natural
frequencies will shift down due to the surface energy. For the first
frequency, the shifting magnitude comes up to 33.5\% of the original
value at $r_0=2.5$nm, and 12.8\% at $r_0=5$nm. Even when $r_0=50$nm,
this ratio can also reach 1.2\%. Therefore, the surface energy
effect is not negligible in the vibration of the silica gel
nanoparticle.

On the other hand, the influence of surface energy on the higher
order frequency is far smaller than that on the first frequency. At
$r_0=2.5$nm, the calculation shows that the shift of third frequency
caused by the surface energy is less than 2\% of this frequency.
Therefore, for the more higher frequency, the surface energy effect
can be ignored.
\begin{figure}[h]
\centering
\includegraphics{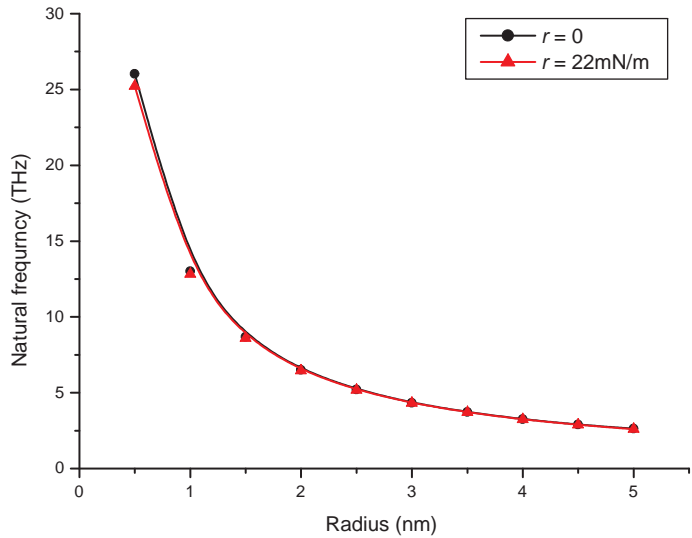}
\caption{The changes of the first natural frequencies with the
radius of the Pd nanoparticle\label{fig:2}}
\end{figure}

Taking $E=123.6$GPa, $\nu=0.39$, $\rho=12023$kg/m$^3$ and
$\gamma=6.0$N/m \cite{30}, we investigate the radial vibration of
the Pd particle with the surface energy effect. The results are
shown in Figure \ref{fig:2}, which depicts the first natural
frequency of the Pd particle changing with radius. From it, it can
be seen that the the influence of surface energy on the first
natural frequency is extremely slight. The calculation shows that
the frequency shifting caused by the surface energy is about 3\% of
this frequency when $r_0=0.5$nm. Therefore, the surface energy
effect can be almost ignored in the vibration of the Pd
nanoparticle.

The results above indicate that, in the vibration, the influence of
surface energy on the silica gel nanoparticle differs from that on
the Pd nanoparticle. From the view of theory, this difference is
determined by relative ratio the elastic modulus to the surface
energy factor. Palladium is a "hard" material. Its surface energy
factor $\gamma$ is far smaller than the elastic modulus in
magnitude. This makes that the influence of $\gamma$ on Eq.
(\ref{30}) is extremely slight so that it can be ignored. However,
silica gel is a soft matter, At the nano-level, the ratio
$\gamma/\lambda r_0>1$. Therefore, the surface energy effect must be
considered in the vibration of the silica gel nanoparticle.

\section{Summary}
\label{sec:4} A vector potential depending only on the field
variables and coordinates is defined. By introducing the divergence
of this vector potential into the Lagrangian, a Lagrangian framework
is developed to incorporate surface energy into the theory of
elasticity. Besides the Euler-Lagrange equation and natural boundary
condition, a new boundary constitutive equation is derived from the
variation of surface area and volume of continuum. The boundary
constitutive equation characterizes the relation, on the boundary
surface of continuum, between the Lagrangian, the vector potential
and the mean curvature of surface. Explicit expression of the vector
potential with respect to the field variable and the mean curvature
is also determined on the boundary surface.

Based on this framework, an elastic model with relevance to surface
energy is established. In this model, a generalized Young-Laplace's
formula is derived consistently. Compared with the existing results,
the generalized Young-Laplace's formula depends not only on surface
energy and surface curvature but also on the normal displacement of
surface. Making use of this model, we investigate the surface energy
effect in the radial vibration of spherical nanoparticle. By
calculation, some conclusions are given as
follows:\\

(1) At the nano-level, the influence of surface energy on the
natural frequencies of a solid particle depends on the relative
difference between elastic modulus and the surface energy factor of
material in magnitude. The surface energy effect can be ignored in
the vibration of hard nanoparticles, but for soft nanoparticles, it
is necessary to take this effect into account.

(2) For a soft nanoparticle, its natural frequencies will shift down
due to the surface energy effect. The smaller the size is, the
stronger the frequencies shifting.

(3) In the vibration of soft nanoparticle, the influence of surface
energy on first natural frequency is far greater than that on the
other frequencies. For the higher order frequencies beyond the third
frequency, the surface energy effect can be ignored.
\section*{Acknowledgements}
The author is very grateful to the reviewer for comment, suggestion
and help! The support of the National Nature Science Foundation of
China through the Grant No. 11172130 is gratefully acknowledged.

\end{document}